\documentclass[aps, showpacs,prb, 10pt, twocolumn]{revtex4}
\usepackage[english]{babel}
\usepackage[latin1]{inputenc}

\usepackage{amssymb}
\usepackage{amsmath}

\usepackage{bm}
\usepackage{graphicx}
\usepackage{epsfig}

\usepackage{dcolumn}
\usepackage{hyperref}

\begin{document}

\preprint{APS/123-QED}

\title{Superconductivity in graphene stacks: from the bilayer to graphite}

\author{Lizardo H. C. M. Nunes}
\email{lizardonunes@ufsj.edu.br}
\author{A. L.  Mota}
\affiliation{Departamento de Ci\^encias Naturais, Universidade Federal de S\~ao Jo\~ao del Rei\\
36301-160 S\~ao Jo\~ao del Rei, MG, Brazil }
\author{E. C. Marino}
\affiliation{Instituto de F\'{\i}sica,  Universidade Federal do Rio de Janeiro,\\
Cx.\,P.\,68528,  Rio de Janeiro-RJ 21941-972, Brazil}
%


\date{\today}

\begin{abstract}
We study the superconducting phase transition, both in a graphene bilayer and in graphite. For that purpose
we derive the mean-field effective potential for a stack of graphene layers presenting hopping between adjacent sheets.
For describing superconductivity, we assume there is an on-site attractive interaction between electrons and determine
the superconducting critical temperature as a function of the chemical potential. This 
displays a dome-shaped curve, in agreement with previous results
for two-dimensional Dirac fermions~\cite{Smith2009,Nunes2005}.
We show that the hopping between adjacent layers increases the critical temperature for small values of the chemical potential.
Finally, we consider a minimal model for graphite~\cite{Pershoguba2010}
and show that the transition temperature is higher than that for the graphene bilayer for small values of chemical potential. This might explain why intrinsic superconductivity is observed in graphite.
\end{abstract}


\maketitle


\section{Introduction}\label{Introduction}

Graphene is a one-atom-thick layer of graphite~\cite{Neto_RMP09}.
The carbon atoms in each layer are arranged in a honeycomb lattice
and the tight-binding energy presents a band structure such that
the valence and conduction bands touch precisely
in the vertices of two inequivalent Dirac cones in the Brillouin zone.
The electronic excitations appearing in the conduction band
have the dispersion relation of a relativistic massless particle
and their properties, accordingly, will be determined by the Dirac equation.
Graphene is believed to be the parent compound of most of the carbon-based systems
and their electric, magnetic and elastic properties all originate from the properties of graphene.

Interestingly, several carbon-based compounds present superconductivity.
For instance, the graphite intercalated compounds (GIC)~\cite{Csanyi2005}
which consists of graphene sheets alternated by alkali layers,
mainly acting as charge reservoirs,
becomes superconducting with the transition temperature
ranging from below 1K for KC$_{ 8 } $ to 11.5 K for CaC $_{ 6 } $~\cite{Weller2005,Emery2005,Belash2002,Hannay1965};
some fullerides present critical temperatures as high as 33 K
as applied pressure or the chemical composition increases the lattice parameter~\cite{Gunnarson1997};
and there are reports of room temperature local superconductivity
within isolated ``grains'' in highly oriented pyrolitic graphite (HOPG)~\cite{Kopolevich2007}
and also with critical temperature $ T_{ c } \sim $ 25 K in thin samples~\cite{Esquinazi2008}.
Moreover, a fully saturated hydrocarbon derived from a single graphene sheet,
called graphane, is predicted to be a high-temperature electron-phonon superconductor
exhibiting a critical temperature of above 90 K~\cite{Savini2010}.

Despite the fact that theoretical conjectures have been proposed
as possible candidates to produce superconductivity~\cite{Meng2010,Pathak2010,Kopnin2008,Baskaran2002,Jiang2008,Black-Schaffer2007,Roy2010},
intrinsic superconductivity has never been observed in graphene,
but it could be only induced by proximity effects,
where a superconducting current propagated
through a superconductor-normal-superconductor (SNS) Josephson junction,
with graphene as the N region~\cite{Heersche2007}.

Nevertheless, the stability of the superconducting phase has been investigated
in graphene~\cite{Pellegrino2010,Khveshchenko2009,Honerkamp2008}
and the symmetry of the order parameter in the honeycomb lattice was identified;
if there is an on-site net attractive interaction between electrons in the honeycomb lattice,
the usual $ s $-wave singlet pairing is favoured~\cite{Zhao2007}.
As nearest-neighbours attraction are taken into account,
an exotic combination of $ s $-wave and $ p $-wave superconducting order parameters is possible~\cite{Uchoa2007}.
In the context of the $ t $-$ J $-$ U $ model, $  f $-wave triplet-pairing and $ d + i d $ singlet-pairing instabilities
are found to emerge away from half-filling~\cite{Honerkamp2008}.

Previously, some of us have investigated
the phase diagram of a quasi-two-dimensional interacting Dirac electrons system
forming Cooper pairs in the singlet state,
which is a suitable model to describe a stack of uncoupled superconducting graphene sheets,
and we have found a quantum critical point connecting
the normal and superconducting phases at a certain critical coupling~\cite{Marino2006}.
If low magnetic fields are applied to the system,
we have found a critical field as a function of the superconducting interaction~\cite{Marino2007}.

In those previous investigations, the variation of the chemical potential was not taken into account;
however, applying a bias voltage, the carrier density of graphene can be controlled by electric field effect.
Therefore, in the present paper we investigate the effect of the chemical potential as a free parameter of our model
and we also consider the effect of the out-of-plane hopping between adjacent graphene sheets.
In order to describe graphite, we consider the minimal model with the electron tunneling
between the nearest sites in the plane and out of the plane.
We have found that the superconducting critical temperature
is enhanced at small values of the chemical potential
for graphite when compared to the values predicted by us for graphene bilayer,
what might explain why intrinsic superconductivity has been observed in HOPG.

The paper is organized as follows:
In Sec.~\ref{TheModel} we present the model Hamiltonian for the graphene bilayer,
the dispersion relation is calculated and the effective potential (free energy) is derived.
In Sec.~\ref{T=0}  the superconducting phase diagram at $ T = 0 $
is obtained analyzing the minima conditions for the effective potential
for several values of the interaction and the hopping between adjacent graphene sheets.
In Sec.~\ref{Tneq0} we calculate the superconducting critical temperature as a function of the chemical potential
for several values of the hopping parameter between layers.
The results represent an upper bound for the Kosterlitz-Thouless transition.
In Sec.~\ref{graphite} our results for the superconducting phase diagram are extended
for an infinite number of coupled graphene layers considering the electron tunneling amplitudes between
the nearest sites in the plane and out of the plane.
Sec.~\ref{Conclusions} is devoted to the conclusion.

\section{Graphene bilayer}\label{TheModel}

\begin{figure}[ht]
\centerline
{ 
\includegraphics[clip, angle=0, width=0.49\textwidth]{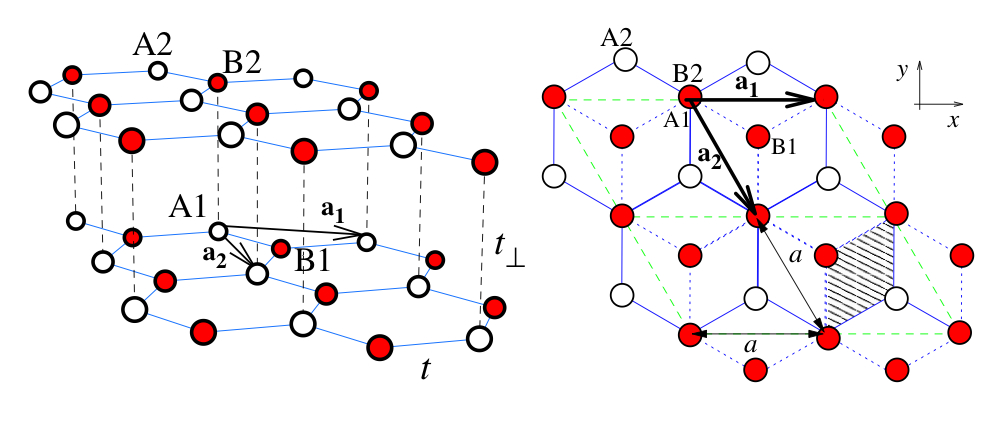}
}
\caption{Lattice structure of two adjacent graphene layers (after \cite{Nuno_Book_2007}).}
\label{FigBilayer}
\end{figure}

Consider a stack of $ N $ graphene layers
with a hopping term between adjacent planes,
where the upper layer has its $ B $ sublattice
on top of sublattice $ A $
of the underlying layer (Bernal stacking),
as can be seen in Fig.\,\ref{FigBilayer}.
The Hamiltonian of each coupled layer
is described by the following \cite{Nuno_Book_2007},
\begin{eqnarray}
H_{ t, l }
& =  &
- \mu \sum_{ {\bf k }, \sigma }
\left[
a^{ \dagger }_{ {\bf k }, \sigma, l } a_{ {\bf k }, \sigma, l }
+
b^{ \dagger }_{ {\bf k }, \sigma, l } b _{ {\bf k }, \sigma, l }
\right]
\nonumber \\
& &
-t \sum_{ {\bf k }, \sigma }
s_{ k }
\left[
a^{ \dagger }_{ {\bf k }, \sigma, l } b_{ {\bf k }, \sigma, l }
+
a^{ \dagger }_{ {\bf k }, \sigma, l+1 } b_{ {\bf k }, \sigma, l+1 }
\right]
+ \mbox{h.c. }
\nonumber \\
& &
-t_{ \bot }
\sum_{ {\bf k }, \sigma }
a^{ \dagger }_{ {\bf k }, \sigma, l } b_{ {\bf k }, \sigma, l+1 }
+ \mbox{h.c. }
\,  ,
\label{EqUnbiasedBilayer}
\end{eqnarray}
where the index $ l = 1, \cdots,  N $ characterizes the different planes
and $ \mu $ is the chemical potential.
The second line in the RHS of the above equation describes
the hopping between electrons of different sublattices
within a graphene sheet,
while the third line describes the hopping between layers.
The hopping parameter is about
$ t \approx 2.8 $ eV
and
$ t_{ \bot } \approx t / 10 $.
The operators
$  a^{ \dagger }_{ i, \sigma, l } =
\sum_{ k }
e^{ i { \bf k } \cdot { \bf r}_{ i } }
\,
a^{ \dagger }_{ {\bf k }, \sigma, l }
$
and
$  b^{ \dagger }_{ i, \sigma, l } =
\sum_{ {\bf k } }
e^{ i { \bf k } \cdot { \bf r}_{ i }  }
\,
b^{ \dagger }_{ {\bf k }, \sigma, l }
$
create, respectively,
an electron on site $ i $
with spin
$ \sigma $
on sublattice $ A $
and
an electron on site $ i $
with spin
$ \sigma $
on sublattice $ B $ of plane $ l $.
In the honeycomb lattice
we have
$ s_{ k } = 1
+ e^{ i { \bf k } \cdot { \bf  a }_{ 1 } }
+ e^{ i { \bf k } \cdot { \bf a }_{  2 } } $,
where $ {\bf a }_{ 1 } = a \hat{ e }_{ x } $ and
$ 2 {\bf a }_{ 2 } = a \left(  \hat{ e }_{ x } - \sqrt{ 3 } \hat{ e }_{ y } \right) $,
as shown in Fig.\,\ref{FigBilayer}.
The lattice parameter is $ a = $ 2.46 \AA for graphene.

We add an on-site attractive interaction
between the electrons within each graphene layer forming
Cooper pairs in the $ s $-wave state.
The interaction term is given by
\begin{eqnarray}
H_{  \mbox{\scriptsize{SC}}, l }
& = &
 - g
\sum_{ {\bf k}, {\bf k}', \sigma}
\left(
a^{\dagger}_{ {\bf k }, \sigma, l }a^{\dagger}_{ - {\bf k }, -\sigma, l}
a_{ - {\bf k }', -\sigma, l }a_{ {\bf k }', \sigma, l }
\right.
\nonumber
\\
& &
+
\left.
b^{\dagger}_{ {\bf k }, \sigma, l }b^{\dagger}_{ - {\bf k }, -\sigma, l }
b_{ - {\bf k }', -\sigma, l }b_{ {\bf k }', \sigma, l }
\right)
\, ,
\label{EqHPairing}
\end{eqnarray}
with $ g > 0 $.
The origin of the interaction
is to be determined by some underlying microscopic theory,
which is not considered here.
However, the symmetry of the gap originated from this interaction
is consistent with the isotropic $ s $-wave symmetry gap observed in some GICs~\cite{Kremer2007}.

Introducing the following Nambu fermion field,
\begin{equation}
\Psi^{ \dagger }_{ { \bf k }, l }
=
\left(
\psi^{ \dagger }_{ { \bf k }, l }, \psi^{ \dagger }_{ { \bf k }, l + 1 }
\right)
\, ,
\label{EqNambu}
\end{equation}
where
\begin{equation}
\psi^{ \dagger }_{ { \bf k }, l }
=
\left(
a^{ \dagger }_{ { \bf k }, \uparrow , l}
\,
b^{ \dagger }_{ { \bf k }, \uparrow , l}
\,
a_{ -{ \bf k }, \downarrow  , l}
\,
b_{ -{ \bf k }, \downarrow  , l}
\right)
\, ,
\label{EqNambu2}
\end{equation}
one can rewrite the combined Hamiltonian
$ H_{ t, l } + H_{ \mbox{\scriptsize{SC}}, l } $
at the mean-field level,
\begin{equation}
H_{  \mbox{\scriptsize{MF}} }
=
\sum_{ \bf k }
\Psi^{ \dagger }_{ { \bf k }, l }
\, \mathcal{A} \,
\Psi_{ { \bf k }, l }
- \frac{ \Delta \Delta^{ * } }{ g }
\,
\label{EqHMF}
\end{equation}
where, by definition, the superconducting order parameter is
\begin{equation}
 - \frac{ \Delta }{ g }
=
\sum_{ \bf k }
\langle
a^{\dagger}_{ {\bf k }, \uparrow, l }a^{\dagger}_{ - {\bf k }, -\downarrow, l}
\rangle
=
\sum_{ \bf k }
\langle
b^{\dagger}_{ {\bf k }, \uparrow, l }b^{\dagger}_{ - {\bf k }, -\downarrow, l}
\rangle
\label{EqDefGap}
\end{equation}
and the 8 $ \times $ 8 matrix $ \mathcal{A} $ in Eq.~(\ref{EqHMF}) is given by
\begin{eqnarray}
\mathcal{A} =
\begin{pmatrix}
 \mathcal{ A }_{ 1 } &  \mathcal{ A }_{ 1 2 } \\
 \mathcal{ A }_{ 2 1 } &  \mathcal{ A }_{ 2 }
\end{pmatrix}
\, ,
\label{EqMatrixA}
\end{eqnarray}
with
\begin{equation}
\mathcal{A}_{ 1 } = \mathcal{A}_{ 2 }
=
\begin{pmatrix}
- \mu                    &  - t s_{ k }          & 0              & \Delta            \\
 - t s^{ * }_{ k }    &  -\mu                 & \Delta       & 0                   \\
 0                        & \Delta^{ * }        & \mu          & t s^{ * }_{ k }  \\
 \Delta^{ * }         &  0                      & t s_{ k }    & \mu                \\
\end{pmatrix}
\label{EqMatrixA1}
\end{equation}
and
\begin{equation}
\mathcal{A}_{ 1 2 } = \mathcal{A}^{ T }_{ 2 1 }
=
\begin{pmatrix}
 0   &  -t_{ \bot }  & 0  & 0               \\
 0   &  0               & 0  & 0               \\
 0   &  0               & 0  & t_{ \bot }  \\
 0   &  0               & 0  & 0               \\
\end{pmatrix}
\, .
\label{EqA12}
\end{equation}

From $ H_{  \mbox{\scriptsize{MF}} } $
in Eq.~(\ref{EqHMF}),
follows the dispersion relation,
\begin{equation}
E_{ k }
=
\pm
\sqrt{
| \Delta |^{2 } + E^{ 2 }_{  \mbox{\scriptsize{BL}} }
}
\, ,
\label{EqDispersion}
\end{equation}
where
\begin{equation}
E_{  \mbox{\scriptsize{BL}} }
=
\pm \sqrt{
t^{2} | s_{ k } |^{ 2 }
+
\left(
\frac{ t_{ \bot } }{ 2 }
\right)^{ 2 }
}
\pm
\frac{ t_{ \bot } }{ 2 } - \mu
\, .
\label{EqDispersionBL}
\end{equation}

\begin{figure}[ht]
\centerline
{ \includegraphics[clip,angle=0, width=0.47\textwidth] {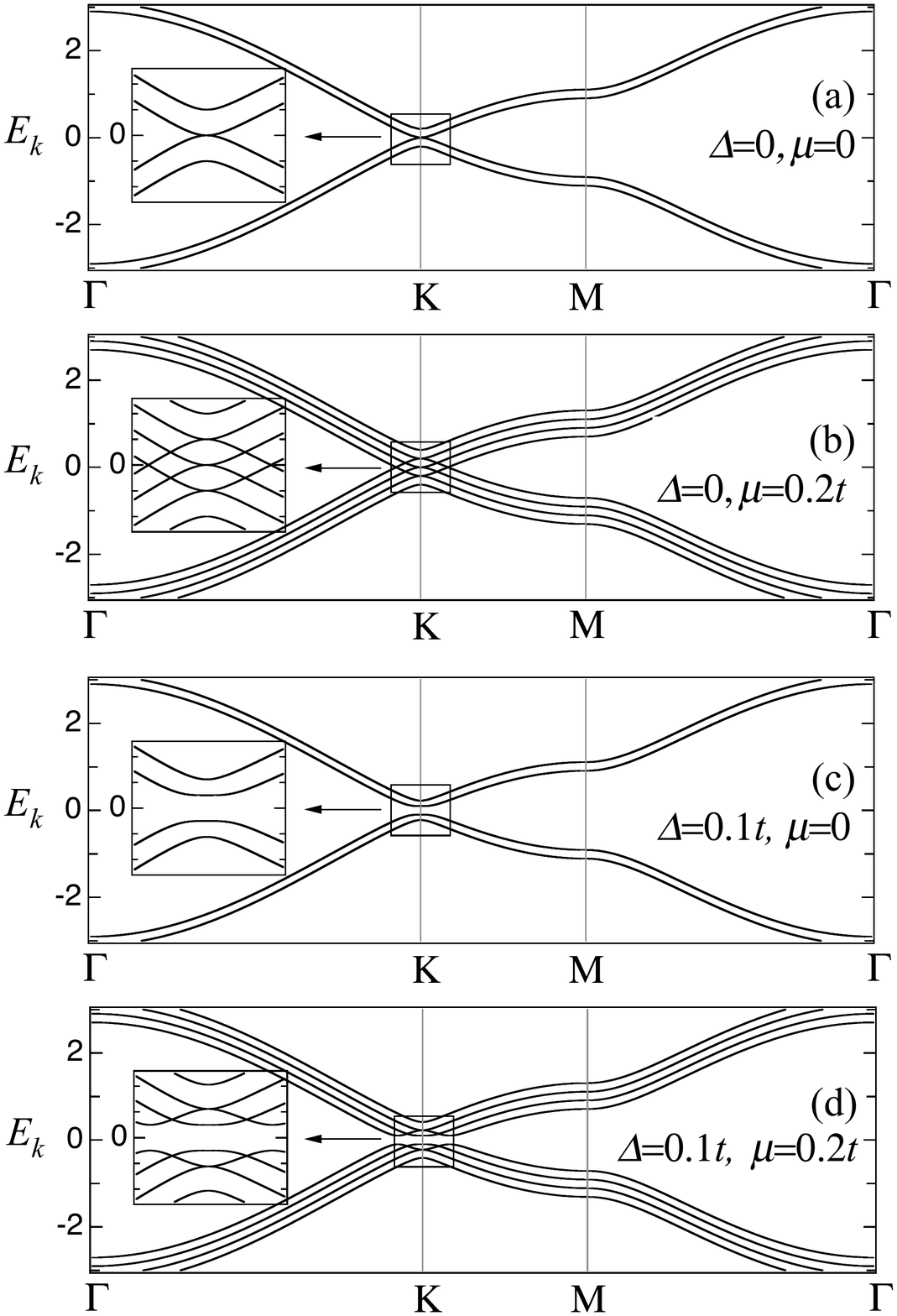} }
\caption{Band structure for
(a) $ \Delta = \mu = 0 $,
(b) $\Delta = 0 $ and $ \mu = 0.2 t $,
(c) $ \Delta = 0.1 t $ and $ \mu = 0 $
(d) $ \Delta = 0.1 t $ and $ \mu = 0.2 t $.
Energy is given in units of  $ t $ and $ t_{ \bot } = 0.2 t $.}
\label{FigBands}
\end{figure}

Let us neglect the hopping term between planes for a moment
and consider only the normal state of the system at the Fermi level,
which means $ t_{ \bot } = \Delta = \mu = 0 $.
In that case,
$ \mathcal{A} $ has eight eigenvalues,
but only two are undistinguished:
$ \pm t \sqrt{ | s_{ \bf k } |^{ 2 } }$,
which is exactly the dispersion relation
of a single layer for a given spin state~\cite{Nuno_Book_2007,Neto_RMP09}.
For $ \Delta \neq 0 $, $\mu \neq 0 $,
we have $ \pm \sqrt{ | \Delta| ^{2 } + \left( \mu \mp t |s_{ k } |  \right)^{2} } $
for each layer, which is the spectrum for the $ s $-wave pairing~\cite{Uchoa2007}.

If we take into account
the hopping term between planes,
in the absence of superconductivity,
we obtain
$ \pm | E_{  \mbox{\scriptsize{BL}} } |$.
In particular, at $ \mu = 0 $,
the four energy bands
along three directions in the first Brillouin zone
for $ \Delta = \mu = 0 $
can be seen in Fig. \ref{FigBands}.a,
which is the same plot shown
for the unbiased graphene bilayer in
\cite{Nuno_Book_2007}.
The eight distinct energy bands
in the normal state are shown in Fig. \ref{FigBands}.b.
(It should be noticed that, for this particular choice of parameters,
the chemical potential is sitting right at the bottom of the upper band.)
As expected,
the system is gapped in the superconducting state and
the four energy bands at $ \mu = 0 $ are shown in Fig \ref{FigBands}.c.
Finally, the eight energy bands for nonzero gap and chemical potential
are shown in Fig. \ref{FigBands}.d.

The graphene dispersion relation has six Dirac points at the corners
of the first Brillouin zone; however, only two of them are non-equivalent.
The continuum limit of our model Hamiltonian is obtained
expanding Eq.~(\ref{EqHMF}) in the vicinity of the Dirac points
$ {\bf K } =  - 4 \pi / 3 a \, \hat{ e }_{ x } $
and
$ {\bf K }' =  4 \pi / 3 a \, \hat{ e }_{ x } $,
\begin{equation}
H^{ \mbox{\scriptsize{CL}} }_{  \mbox{\scriptsize{MF}}, l }
=
\sum_{ \alpha }
\int \frac{ d^{ 2 } k }{ \left( 2 \pi \right)^{ 2 } }
\,
\Psi^{ \dagger }_{ \alpha, l } ( k )
\, \mathcal{A}_{ \alpha }  \,
\Psi_{ \alpha, l } ( k )
- \frac{ \Delta \Delta^{ * } }{ g }
\, ,
\label{EqHCL}
\end{equation}
where $ \alpha = K , K' $
and $ \mathcal{ A }_{ \alpha } $
is obtained
replacing $ t s_{ k } $ by
$ - v_{ \rm{ F } } \left(  k_{ x } - i k_{ y } \right) $
and
$ - v_{ \rm{ F } } \left(  k_{ x } + i k_{ y } \right) $
in Eq.\,(\ref{EqMatrixA1})
for $ K  $ and $ K' $ respectively,
with $ \hbar = 1 $ and  $ v_{ \rm{ F } } = \sqrt{ 3 }  t a / 2 $.

The partition function in the complex time representation is written as
\begin{eqnarray}
\mathcal{ Z }
& = &
\frac{ 1 }{ \mathcal{ Z }_{ 0 } }
\int
\mathcal{ D } \Psi^{ * } \mathcal{ D }  \Psi
\exp
\left\{
\sum_{ l = 1 }^{ N }
\int_{ 0 }^{ \beta } d \tau
\,
L^{ \mbox{\scriptsize{CL}} }_{  \mbox{\scriptsize{MF}}, l }
\right\}
\, ,
\label{EqZ}
\end{eqnarray}
where $\mathcal{ Z }_0 $ is the vacuum functional,
$ \beta = 1 / k_{ B } T $ ($ k_{B }$ is the Boltzmann constant)
and
\begin{equation}
L^{ \mbox{\scriptsize{CL}} }_{  \mbox{\scriptsize{MF}}, l }
=
\sum_{ \alpha, \sigma }
\,
\int \frac{ d^{ 2 } k }{ \left( 2 \pi \right)^{ 2 } }
\,
\left(
 \psi^{ \dagger }_{ \alpha, \sigma, l}  i \partial_{ \tau } \psi_{ \alpha, \sigma, l}
-
H^{ \mbox{\scriptsize{CL}} }_{  \mbox{\scriptsize{MF}}, l }
\right)
\, ,
\label{EqLCL}
\end{equation}
with
$ \psi^{ \dagger }_{ \alpha, \sigma, l}
= \left[
a^{ \dagger }_{ \alpha, \sigma, l} ( k ),
b^{ \dagger }_{\alpha, \sigma, l } ( k )
\right] $ representing the spinorial fields appearing
in the continuum limit of the tight-biding graphene Hamiltonian density.

Integrating over the fermion fields,
we get that the partition function is proportional to
\begin{equation}
\left(
\frac{
\mbox{det} \, \mathcal{ A }'_{ \alpha, n }
}{
\mbox{det} \, \mathcal{ A }'_{ \alpha, n } [ \Delta = 0 ]
}
\right)^{ 2 N }
\, ,
\label{EqFermionIntegral}
\end{equation}
where
$ \mathcal{ A }'_{ \alpha, n } =  - i \omega_{ n } { \bf 1 } + \mathcal{ A }_{ \alpha } $
is a function of the  Matsubara frequencies for fermions, $\omega_n = (2 n+1)\pi T$,
and $ { \bf 1} $ is the 8 $ \times $ 8 unity matrix,
with $ k_{ B } = 1 $ hereafter for the sake of simplicity.

Finally, redefining the coupling $ g = \lambda / N $,
the ``effective potential'' per bilayer
for each Dirac point will be
\begin{eqnarray}
V_{ \rm eff}
& = &
2 \frac{ \Delta \Delta^{*} }{ \lambda }
-
\frac{ 1 }{ \beta } \sum_{ n }
\left[
\int \frac{ d^{ 2 } k }{ \left( 2 \pi \right)^{ 2 } }
\right.
\nonumber \\
& &
\; \; \; \; \; \; \; \; \; \; \; \; \; \; \; \; \; \; \; \; \; \; \;
\left.
\ln
\left(
\frac{
\mbox{det} \, \mathcal{ A }'_{ K, n }
}{
\mbox{det} \, \mathcal{ A }'_{ K, n } [ \Delta = 0 ]
}
\right)
\right]
.
\label{EqVeff}
\end{eqnarray}
This would be the leading order in a $ 1 / N $ expansion and would be the exact result for $ N \rightarrow \infty $.

In the next section,
we analyze the conditions for the appearance of superconductivity at zero temperature
provided by our mean-field effective potential.

\subsection{The superconducting instabilities at $ T  = 0 $}
\label{T=0}

We shall study the minima of the effective potential.
The occurrence of superconductivity
corresponds to the existence of nonzero solutions of the order parameter,
which minimizes the effective potential.

Taking the derivative of $ V_{\rm eff} $
with respect to the order parameter and summing over the Matsubara frequencies,
we obtain
\begin{eqnarray}
V'_{\rm eff}( T )
& = &
\Delta^{ * }
 \left[
\frac{2}{\lambda }
-
\frac{ 1 }{ 2 }
\sum_{ j = 1 }^{ 4 }
\int \frac{ d^{ 2 } k }{ \left( 2 \pi \right)^{ 2 } }
\right.
\nonumber \\
& &
\hspace{-0.3cm}
\left.
\frac{ 1  } { \sqrt{ | \Delta |^{2 } + \xi_{ j }^{ 2 }  } }
\tanh
	\left(
	         \frac{ \beta }{ 2 } \sqrt{ | \Delta |^{2 } + \xi_{ j }^{ 2 }  }
	\right)
\right]
\, ,
\label{EqV'eff}
\end{eqnarray}
where
\begin{equation}
\xi_{ j }
=
\pm
\sqrt{
v_{ \rm F }^{2} k^{ 2 }
+
\left(
\frac{ t_{ \bot } }{ 2 }
\right)^{ 2 }
}
\pm
\frac{ t_{ \bot } }{ 2 } - \mu
\, .
\label{EqXi}
\end{equation}

The nonzero solutions for $ | \Delta | $ are given
equalizing to zero the expression within brackets
in the Eq.~(\ref{EqV'eff}) above,
which provides a self-consistent gap equation.
In particular, at zero temperature, we get
\begin{equation}
V'_{\rm eff}( 0 )
=
\Delta^{ * }
 \left[
\frac{2}{\lambda }
-
\frac{ 1 }{ 2 }
\sum_{ j = 1}^{ 4 }
\int \frac{ d^{ 2 } k }{ \left( 2 \pi \right)^{ 2 } } \,
\frac{ 1  } { \sqrt{ | \Delta |^{2 } + \xi_{ j }^{ 2 }  } }
\right]
\, .
\label{EqV'effT0a}
\end{equation}

Introducing a large momentum cutoff
$\Lambda/v_{ \rm{F}}$,
we can integrate Eq.~(\ref{EqV'effT0a}) over $ k $,
\begin{eqnarray}
V'_{\rm eff}
& = &
\Delta^{ * }
\left\{
\frac{2}{\lambda }
-
\frac{ 1 }{ 2 \alpha }
\sum_{ a, b }
\left[
\sqrt{ | \Delta |^{2 } + \xi^{ 2 }_{ a b } }
- \sqrt{ | \Delta |^{2 } + \epsilon^{ 2 }_{ a b } }
\right.
\right.
\nonumber \\
& + &
\left.
\left.
\left( \mu -  b \frac{t_{ \bot } }{ 2 } \right)
\ln
	\left(
	         \frac{ \sqrt{ | \Delta |^{2 } + \xi^{ 2 }_{ a b } }  + \xi_{ a b }  }
	                 { \sqrt{ | \Delta |^{2 } + \epsilon^{ 2 }_{ a b } + \epsilon_{ a b } }     }
	\right)
\right]
\right\}
\, ,
\label{EqV'effT0}
\end{eqnarray}
where
$ \alpha = 2  \pi  v^{ 2 }_{ \rm F } $,
$ a, b = \pm 1$,
\begin{equation}
\xi_{ ab }
=
a \sqrt{ \Lambda^{ 2 } + \left( \frac{ t_{ \bot }}{ 2 } \right)^{ 2 } }
+ b \frac{ t_{ \bot }}{ 2 } - \mu
\, ,
\label{EqXiab}
\end{equation}
and $ \epsilon_{ a b} = \xi_{ a b }( \Lambda = 0 )$.

In particular, for $ \mu =  t_{ \bot } = 0 $, the nonzero solutions for the superconducting gap are \cite{Marino2006}
\begin{equation}
\Delta = \frac{ \alpha \lambda  }{ 2 }
\left(
\frac{ \Lambda^{ 2 }   }{ \alpha^{ 2 }  } -  \frac{ 1 }{ \lambda^{ 2 } }
\right)
\, ,
\label{EqDelta00}
\end{equation}
for $ \lambda > \alpha / \Lambda $,
what establishes quantum critical point for the onset of superconductivity in the system
at the critical coupling $ \lambda_c = \alpha / \Lambda $.

Evidencing $ \Lambda $ in (\ref{EqV'effT0}), it can be reexpressed as
\begin{eqnarray}
V'_{\rm eff}
& = &
\frac{ \Delta^{ * } }{ \tilde{ \alpha } }
\left\{
\frac{2}{\lambda'}
-
\frac{ 1 }{ 2 }
\sum_{ a, b }
\left[
\sqrt{ | \tilde{ \Delta } |^{2 } + \tilde{ \xi }^{ 2 }_{ a b } }
- \sqrt{ | \tilde{ \Delta } |^{2 } + \tilde{ \epsilon }^{ 2 }_{ a b } }
\right.
\right.
\nonumber \\
& + &
\left.
\left.
\left( \tilde{ \mu } -  b \frac{ \tilde{t}_{ \bot } }{ 2 } \right)
\ln
	\left(
	         \frac{ \sqrt{ | \tilde{ \Delta } |^{2 } + \tilde{ \xi }^{ 2 }_{ a b } }  + \tilde{ \xi }_{ a b }  }
	                 { \sqrt{ | \tilde{ \Delta } |^{2 } + \tilde{ \epsilon }^{ 2 }_{ a b } + \tilde{ \epsilon }_{ a b } }     }
	\right)
\right]
\right\}
\, ,
\label{EqV'effT02}
\end{eqnarray}
where $ \lambda' = \lambda / \lambda_{ c }$ and the tilde indicates that the quantity is divided by $ \Lambda $.
Since all the nonzero solutions for the gap are given by the expression between the curly brackets above, 
and given that it does not depend on any experimental data,
our results are suitable to describe any planar Dirac fermion system 
with a hopping between adjacent sheets, 
assuming that $ \Lambda $ is the single free parameter of our model.
Therefore, in the following we present our numerical results for $ \Delta $ in terms of the parameter $ \Lambda $.

Notice that the cutoff is always provided by the lattice in condensed matter systems.
Indeed, we have $ \Lambda = 2 \pi \hbar v_{ \rm F }/ a $ as an upper bound for the energy cutoff,
and since $ a $ is the smallest distance scale, $ \Lambda $ becomes a natural high-energy cutoff.
In fact, this frequently happens in condensed matter.
A familiar example in the case of conventional, phonon mediated superconductivity, 
is the Debye frequency (energy) a natural cutoff that emerges in BCS theory.
Moreover, we also constrain ourselves to positive values of the chemical potential up to $ \mu / \Lambda = 0.9 $,
given the half bandwidth of $ \Lambda $.

\begin{figure}[ht]
\centerline
{
\includegraphics[clip,angle=0, width=0.47\textwidth]{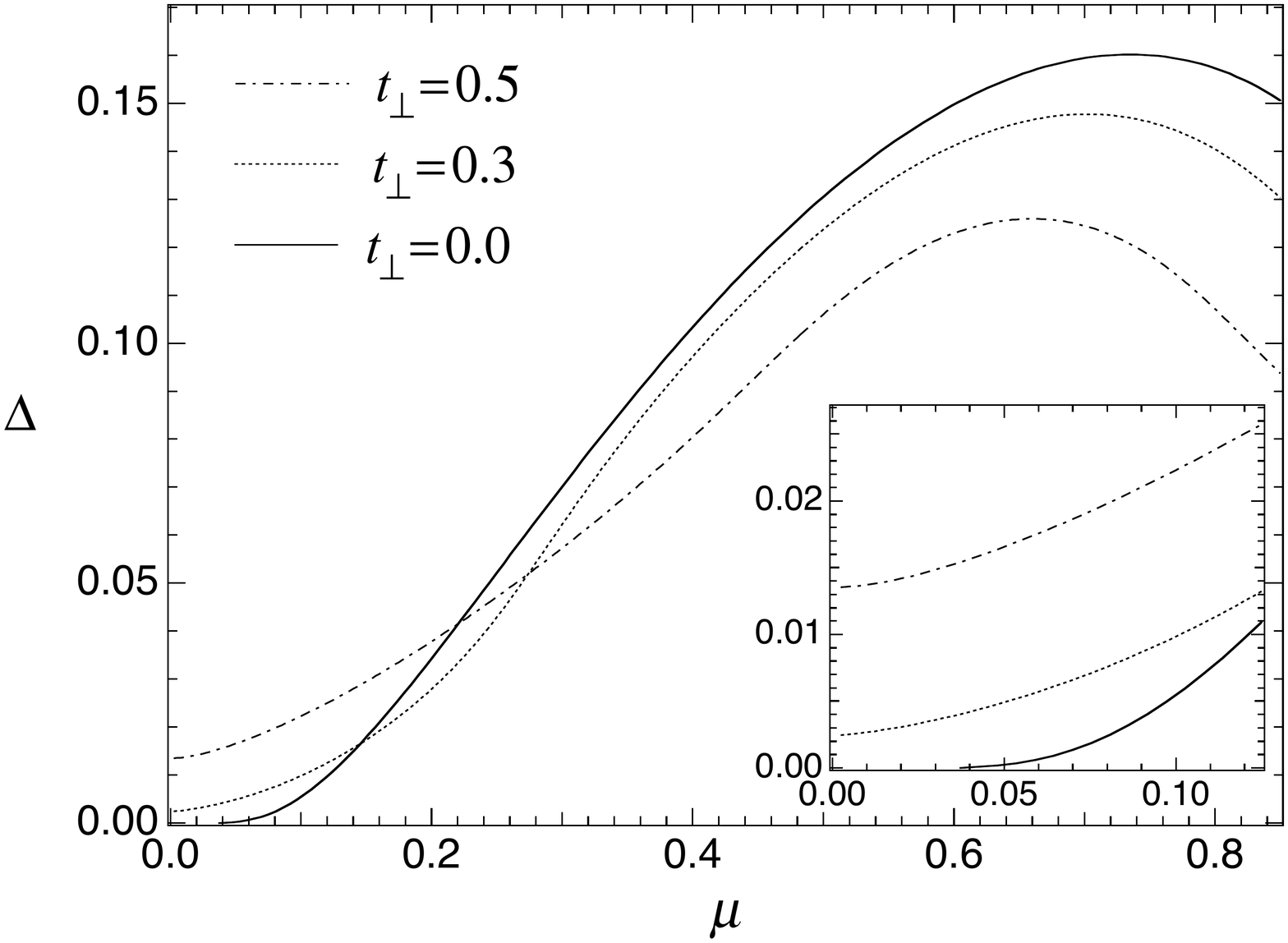}
}
\caption{The superconducting gap as a function of the chemical potential for $ t_{ \bot } = 0, 0.3$ and $ 0.5 $.
The inset shows the same plot for a smaller range of the chemical potential.
$  \lambda / \lambda_{ c }  = 0.8 $ and all the other quantities are given in unities of $ \Lambda $.}
\label{FigGap0_X_mu}
\end{figure}

The case for $ t_{ \bot } = 0 $ and finite $ \mu $ with different values of interaction coupling
has been exhaustively investigated by some of us~\cite{Nunes2010}.
The plots of the superconducting gap as a function of $ \mu $
for $  \lambda / \lambda_{ c }  = 0.8 $
are shown in Fig.~\ref{FigGap0_X_mu}.
Starting at $ \mu / \Lambda  = 0 $, the system is in the normal state,
since $ \lambda < \lambda_{ c } $.
As $  \mu / \Lambda  $ increases, $ \Delta_0 / \Lambda $ displays a dome-shaped plot:
the system asymptotically becomes superconducting
up to a maximum value at an optimal chemical potential
and decreases as $ \mu / \Lambda $ increases even further.

Notice that the system is not quantum critical but, in fact,
the curve vanishes exponentially as $ \mu \rightarrow 0 $,
hence, superconductivity persists down to $ \mu = 0 $.

Our results are consistent with~\cite{Fukushima2007}, which also obtains a
dome-shaped plot of $ \Delta $ for relativistic interacting particles,
as can be seen in Fig.~1 of their paper
(choice of parameters I, referred as the {\it weak-coupling case}).

An interesting result is obtained as we increase the value of $ t_{ \bot } $,
as can bee seen in the inset of Fig.~\ref{FigGap0_X_mu}.
For small values of the chemical potential,
as the out of the plane hopping between layers increases,
we see that the superconducting gap also increases for the same value of the chemical potential.
Indeed, even for $ \mu = 0 $, for which there is no superconducting gap when $ t_{ \bot } = 0 $,
given that $ \lambda < \lambda_{ c } $, there is a nonzero value of $ \Delta $
for $ t_{ \bot } / \Lambda = 0.3 $ or $ 0.5 $,
indicating that the system is in the superconducting state.
Therefore, the hopping between layers favors the appearance of superconductivity.

As shall be seen in the next section,
since the energy gap and the superconducting critical temperature
tend to be proportional quantities,
this result also shows that the superconducting critical temperature increases
as the hopping between layers increases for small values of the chemical potential.

\subsection{Superconducting phase at finite temperatures}\label{Tneq0}

In this section we calculate the superconducting phase diagram for finite temperatures.
{\it A priori}, the nonzero solutions for $\Delta$ are supposed to hold
only in the $ N \rightarrow \infty $ limit at a finite temperature,
because, otherwise, they are ruled out by the Coleman-Mermin-Wagner-Hohenberg theorem~\cite{mw}.
This limit corresponds to a physical situation where the three-dimensionality
of the system is explicitly taken into account.
For finite values of $ N $ and $ T \neq 0 $,
there is an underlying Berezinskii-Kosterlitz-Thouless (BKT) transition~\cite{kt},
below which phase coherence is found for a nonzero $ \Delta $.
The actual superconducting transition occurs at $T_{BKT} \leq T_c $.
However, it can be shown that $T_{BKT} \stackrel{N\rightarrow\infty}{\longrightarrow} T_c$~\cite{babaev}.
This clearly indicates that,
in spite of the fact that we may have a nonzero superconducting gap at $ T = T_{ c } $,
only in a really three-dimensional system we will have phase coherence developing at the same temperature
that the modulus of the order parameter becomes nonzero, as determined by the gap equation.
Therefore, $ T_{ c } $ calculated in this section may be regarded as a mean-field upper bound critical temperature for the KT transition,
which sets the actual temperature for the appearance of superconductivity in the $ N \rightarrow \infty $ limit.

We start considering the gap equation provided by the Eq.~(\ref{EqV'eff}).
Making the change of variables $ x = ( v_{ F } k )^{ 2 } $, we get
\begin{equation}
\frac{ 1 }{\lambda }
-
\frac{ 1 }{ 8 \alpha }
\sum_{ a, b }
\int_{ 0 }^{ \Lambda^{ 2 } } dx
\;
\frac{ 1  } { E_{ a b}( x ) }
\tanh
	\left[
	         \frac{ E_{ a b}( x )  }{ 2 T }
	\right]
= 0
\, ,
\label{EqGapEquationTneq0}
\end{equation}
where
\begin{equation}
E_{ a b}( x )
\equiv
\sqrt{ | \Delta |^{2 } + \xi_{ a b }^{ 2 }( x )  }
\, ,
\label{EqEab}
\end{equation}
with $ \xi_{ a b }( x ) $ given by Eq.~(\ref{EqXiab}), replacing $ \Lambda^{ 2 } $ for $ x $.
From Eq.~(\ref{EqGapEquationTneq0}),
we calculate the superconducting critical temperature $ T_{ c } $
making $ \Delta =  0 $ at $ T = T_{ c } $ in the above expression.

\begin{figure}[ht]
\centerline
{
\includegraphics[clip,angle=0, width=0.47\textwidth]{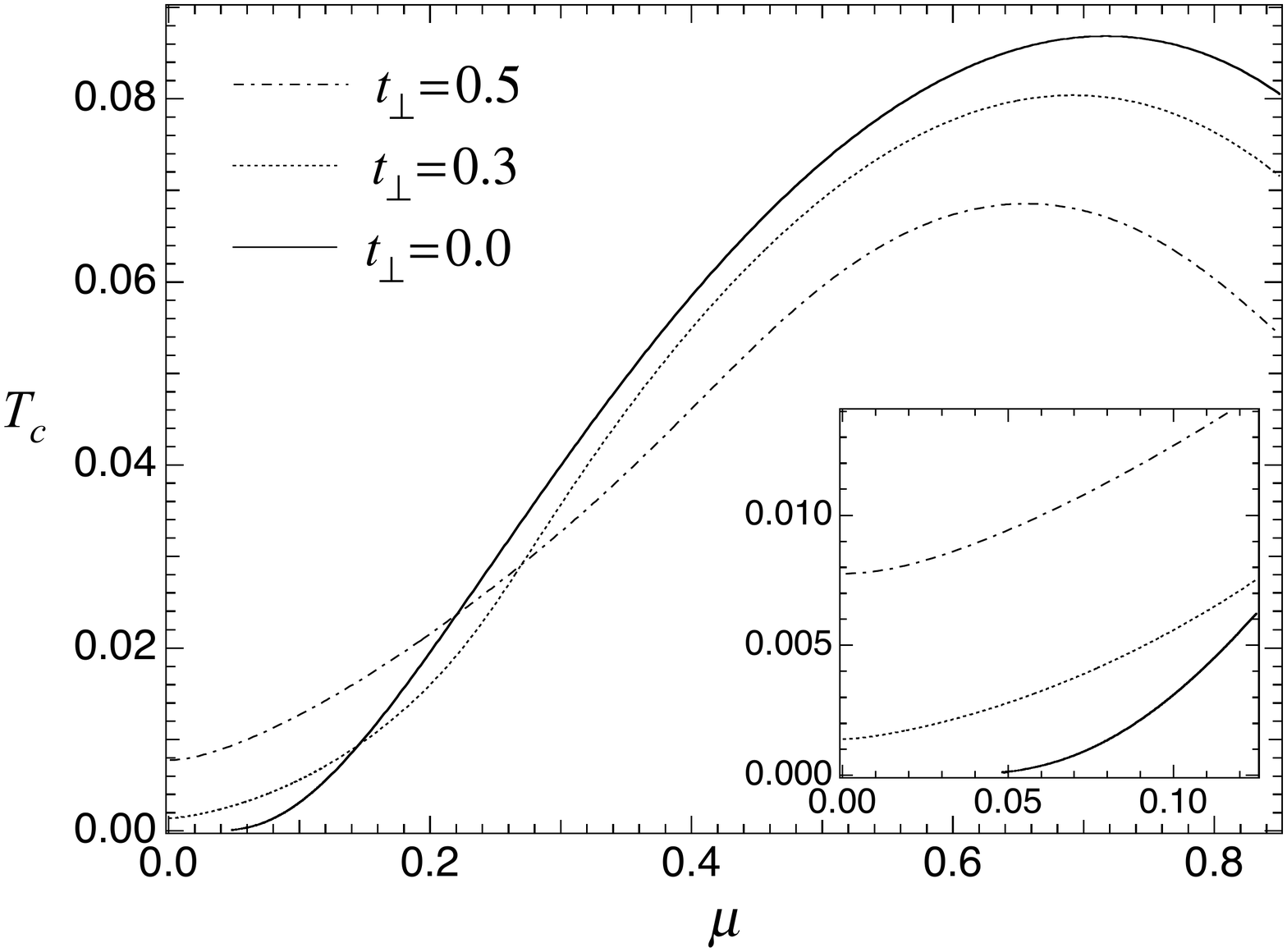}
}
\caption{The superconducting critical temperature as a function of the chemical potential for $ t_{ \bot } = 0, 0.3 $ and $ 0.5 $.
The inset shows the same plot for a smaller range of the chemical potential.
$ \lambda / \lambda_{ c }  = 0.8 $ and all the other quantities are given in unities of $ \Lambda $.}
\label{FigTc_X_mu}
\end{figure}

In Section \ref{T=0} we have found a dome-shaped superconducting gap as a function of the chemical potential
for $ \lambda < 1 $ and several values of $ t_{ \bot } $.
Since the energy gap and the superconducting critical temperature $ T_{ c }$ are proportional,
we expect to find a dome-shaped plot for $ T_{ c } $ as a function of $ \mu $ as well.
In fact, as can be seen in Fig.~\ref{FigTc_X_mu},
our numerical results for the superconducting critical temperature presents the characteristic dome
experimentally observed in several compounds, like 1111 pnictides and cuprate superconductors.

A dome-like structure of the superconducting phase for two-dimensional Dirac fermions
has been previously obtained in \cite{Smith2009},
where the superconducting critical temperature also presents a dome at intermediate filling fractions,
surrounded by the normal phase for fillings close to unity or zero,
which is consistent to our results.
Also, a dome for $ T_{ c } $ as a function of hole concentrations
has been previously obtained in a brief letter by some of us~\cite{Nunes2005}
for a relativistic version of the spin-fermion Hamiltonian~\cite{Kampf1994},
used to describe the Cu-O planes in the cuprates.
Those results and the phase diagram presently calculated
suggest that Dirac fermions may play a relevant role in the description of cuprates and iron pnictides.

Indeed, it has been shown that Dirac points appear in the intersection of the nodes of the d-wave superconducting gap
and the 2D-Fermi surface in the high-Tc cuprate superconductors
and the low-energy excitations will correspond exclusively to these points~\cite{cuprates}.
Also, it has been experimentally found that the iron pnictides~\cite{pnic1, pnic2}
also present electronic excitations whose properties are governed by the Dirac equation.
Theoretical results also support the existence of Dirac electrons in the pnictides~\cite{Direlpnic,Direlpnic1}.

As also suggested in the former section,
the inset of Fig.~\ref{FigGap0_X_mu} shows that,
for a small value of $ \mu $,
$ T_{ c } $ increases as the $ t_{ \bot } $ parameter is increased,
indicating that the hopping between layers favors the appearance of superconductivity in the system.

As shall be seen in the next section, the same feature is observed
as we take into account the first-neighbors out of the plane
hopping between adjacent layers, which is the case for graphite.

\section{Graphite}
\label{graphite}

In this section, we calculate the superconducting phase diagram of many coupled graphene layers
for a finite chemical potential.
To simplify the problem, we consider only the minimal model
where only the electron tunneling amplitudes between the nearest sites in the plane $ t $
and out of the plane $ t_{ \bot } $ are regarded.
The same approach was employed in~\cite{Pershoguba2010},
as briefly explained below.

Consider a Bernal-stacked graphene bilayer described
by the following Hamiltonian in the vicinity of each non-equivalent Dirac point~\cite{Nuno_Book_2007},
\begin{equation}
H_{  \mbox{\scriptsize{BL}} } = \sum_{ {\bf k}, \sigma } \Phi^{ \dagger }_{ {\bf k}, \sigma }  \mathcal{ B }_{\bf k}  \Phi_{ {\bf k}, \sigma }
\, ,
\label{EqHBilayer}
\end{equation}
where the above 4 $ \times $ 4 matrix $  \mathcal{ B }_{\bf k}  $ is given by
\begin{eqnarray}
 \mathcal{ B }_{\bf k}
 =
\begin{pmatrix}
 v_{ F } \, {\bf k } \cdot \vec{\sigma }  &  \mathcal{ B }_{ 1 2 } \\
 \mathcal{ B }_{ 2 1 } &  v_{ F } \, {\bf k } \cdot \vec{\sigma }
\end{pmatrix}
\, ,
\label{EqMatrixB}
\end{eqnarray}
the vector $ \vec{ \sigma } = \left( \sigma_{ x }, \sigma_{ y } \right) $
is written in terms of the well-known Pauli matrices, the matrix $ \mathcal{ B }_{ 1 2 } $ is
\begin{equation}
\mathcal{B}_{ 1 2 } = \mathcal{B}^{ T }_{ 2 1 }
=
\begin{pmatrix}
 0   &  t_{ \bot }  \\
 0   &  0
 \end{pmatrix}
\, ,
\label{EqB12}
\end{equation}
and $ \Phi^{ \dagger }_{ {\bf k }, \sigma }  =  \left( \phi^{\dagger}_{ {\bf k }, \sigma, 1} ,  \phi^{\dagger}_{ {\bf k }, \sigma, 2}  \right) $,
with $ \phi^{\dagger}_{ {\bf k }, \sigma, j} =  \left(  a^{ \dagger }_{ { \bf k }, \sigma , j}  \,  b^{ \dagger }_{ { \bf k }, \sigma , j} \right) $,
$ j = 1, 2 $ denotes the layer index.

The model Hamiltonian for graphite is assumed to be described as an infinite number of graphene layers
coupled by the hopping between adjacent sheets. Therefore, introducing the operator
\begin{equation}
\tilde{ \Phi }^{ \dagger }_{ {\bf k }, \sigma  }
=
\left( \cdots \,
\phi^{\dagger}_{ {\bf k }, \sigma, l -1}
\,
\phi^{\dagger}_{ {\bf k }, \sigma, l }
\,
\phi^{\dagger}_{ {\bf k }, \sigma, l + 1}
\, \cdots \right)
\, ,
\label{EqPhiGraphite}
\end{equation}
the Hamiltonian becomes
\begin{equation}
H_{  \mbox{\scriptsize{Gr}} }
=
\sum_{ {\bf k}, \sigma } \tilde{ \Phi }^{ \dagger }_{ {\bf k}, \sigma }  \mathcal{ C }_{\bf k}  \tilde{ \Phi }_{ {\bf k}, \sigma }
\, ,
\label{EqHGrafite}
\end{equation}
where
\begin{eqnarray}
\mathcal{ C }_{\bf k}
 =
\begin{pmatrix}
\ddots &                                                          &                                                          &                                                &            \\
 &  v_{ F } \, {\bf k } \cdot \vec{\sigma } &  \mathcal{ B }_{ 1 2 }                        &                                                          &            \\
 &  \mathcal{ B }_{ 2 1 }                        &  v_{ F } \, {\bf k } \cdot \vec{\sigma } & \mathcal{ B }_{ 1 2 }                         &            \\
 &                                                          &  \mathcal{ B }_{ 2 1 }                        &  v_{ F } \, {\bf k } \cdot \vec{\sigma } &            \\
 &                                                          &                                                          &                                                          & \ddots \\
\end{pmatrix}
\, .
\label{EqMatrixC}
\end{eqnarray}

Introducing the momentum $ k_{ z } $ in the $ z $ direction,
it is possible to re-express the Hamiltonian for graphite
taking the first-neighbors hopping between adjacent layers in the momentum representation,
which is written in terms of a 4 $ \times $ 4 matrix similar to $ \mathcal{ B }_{\bf k } $
in Eq.~(\ref{EqMatrixB})~\cite{Pershoguba2010},
\begin{equation}
H_{  \mbox{\scriptsize{Gr}} } = \sum_{ {\bf k}, k_{ z }, \sigma } \Phi^{ \dagger }_{ {\bf k}, , k_{ z }, \sigma }  \mathcal{ D }_{ {\bf k}, k_{ z } }  \Phi_{ {\bf k}, , k_{ z }, \sigma }
\, ,
\label{EqHGraphite2}
\end{equation}
where
\begin{eqnarray}
\mathcal{ D }_{ {\bf k}, k_{ z } }
 =
\begin{pmatrix}
 v_{ F } \, {\bf k } \cdot \vec{\sigma }  &  2 \, \mathcal{ B }_{ 1 2 } \cos k_{ z } d \\
2 \, \mathcal{ B }_{ 2 1 } \cos k_{ z } d &  v_{ F } \, {\bf k } \cdot \vec{\sigma }
\end{pmatrix}
\, ,
\label{EqMatrixD}
\end{eqnarray}
and $ d $ is the distance between layers.

For this minimal model, the dispersion relation is given by
\begin{equation}
E_{  \mbox{\scriptsize{Gr}} }
=
\pm \sqrt{ | v_{  \mbox{\scriptsize{F}} } \, {\bf k }  |^{ 2 }
+
\left(
{ t_{ \bot } \cos k_{ z } d }
\right)^{ 2 }
}
\pm
{ t_{ \bot } \cos k_{ z } d }
\,
\label{EqDispersionGraphite}
\end{equation}
and, for $ k_{ z } d = \pi / 2 $, we recover the Dirac-type dispersion found in graphene.

\begin{figure}[ht]
\centerline
{
\includegraphics[clip,angle=0, width=0.4\textwidth]{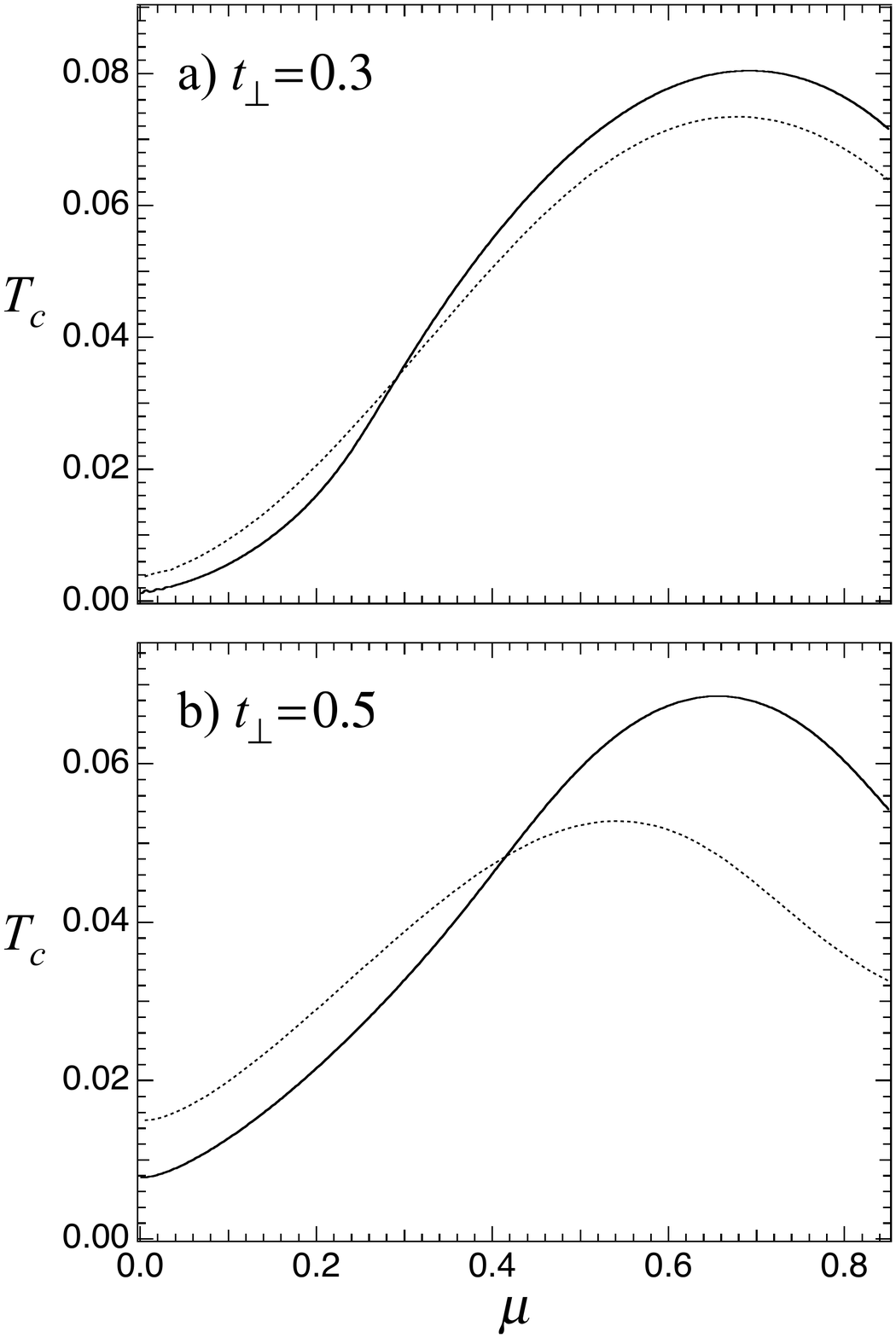}
}
\caption{The superconducting critical temperature as a function of the chemical potential for
 (a) $ t_{ \bot } = 0.3 $ and (b) $ t_{ \bot } = 0.5 $ for both graphene bilayer (solid line) and graphite (dotted line). 
 The inset in each pane shows the same plot for a smaller range of the chemical potential.
$ \lambda / \lambda_{ c }  = 0.8 $ and all the other quantities are given in unities of $ \Lambda $.}
\label{FigTc_X_mu_Graphite}
\end{figure}

Taking into account the attractive interaction forming Cooper pairs
within each graphene layer,
as seen in Eq.~(\ref{EqHPairing}),
and introducing the operator
\begin{equation}
\tilde{ \Psi}^{ \dagger }_{ {\bf k }, \sigma  }
=
\left( \cdots \,
\psi^{\dagger}_{ {\bf k }, \sigma, l -1}
\,
\psi^{\dagger}_{ {\bf k }, \sigma, l }
\,
\psi^{\dagger}_{ {\bf k }, \sigma, l + 1}
\, \cdots \right)
\, ,
\label{EqPsiGraphiteSC}
\end{equation}
where $ \psi^{\dagger}_{ {\bf k }, \sigma, l }  $ is given by Eq.~(\ref{EqNambu2}),
the model Hamiltonian which describes the superconducting graphite in a mean-field approximation becomes
\begin{equation}
H_{  \mbox{\scriptsize{Gr}},  \mbox{\scriptsize{SC}} }
=
\sum_{ {\bf k}, \sigma } \tilde{ \Psi }^{ \dagger }_{ {\bf k}, \sigma }  \mathcal{ E }_{\bf k}  \tilde{ \Psi }_{ {\bf k}, \sigma }
\, ,
\label{EqHSCGrafite}
\end{equation}
where
\begin{eqnarray}
\mathcal{ E }_{\bf k}
 =
\begin{pmatrix}
\ddots &                             &                                        &                                     &            \\
 &   \mathcal{ A }_{ 1  }      &  \mathcal{ A }_{ 1 2 }      &                                     &            \\
 &  \mathcal{ A }_{ 2 1 }     &  \mathcal{ A }_{ 2  }        &  \mathcal{ A }_{ 1 2 }   &            \\
 &                                       &  \mathcal{ A }_{ 2 1 }      &  \mathcal{ A }_{ 1  }     &            \\
 &                                       &                                        &                                    & \ddots  \\
\end{pmatrix}
\, ,
\label{EqMatrixE}
\end{eqnarray}
with $ \mathcal{A}_{ 1 } = \mathcal{A}_{ 2 }  $ and $ \mathcal{A}_{ 1 2 } = \mathcal{A}^{ T }_{ 2 1 } $
given by the Eqs.~(\ref{EqMatrixA1}) and (\ref{EqA12}) respectively.

Accordingly, it is possible to re-express the Hamiltonian for the superconducting graphite in terms
of an 8 $ \times $ 8 matrix, which is similar to $ \mathcal{ A }_{\bf k } $ in Eq.~(\ref{EqMatrixA}),
\begin{equation}
H_{  \mbox{\scriptsize{Gr}} , \mbox{\scriptsize{SC}} }
=
\sum_{ {\bf k}, k_{ z }, \sigma } \Psi^{ \dagger }_{ {\bf k}, , k_{ z }, \sigma }  \mathcal{ F }_{ {\bf k}, k_{ z } }  \Psi_{ {\bf k}, , k_{ z }, \sigma }
\, ,
\label{EqHSCGraphite2}
\end{equation}
where
\begin{eqnarray}
\mathcal{ F }_{ {\bf k}, k_{ z } }
 =
\begin{pmatrix}
\mathcal{ A }_{ 1 }                                &   2 \mathcal{ A }_{ 1 2 }  \cos k_{ z } d  \\
2 \mathcal{ A }_{ 2 1 }  \cos k_{ z } d   &   \mathcal{ A }_{ 2 }                               \\
\end{pmatrix}
\label{EqMatrixF}
\end{eqnarray}
and the dispersion is given by the 8 eigenvalues
\begin{equation}
E_{ \pm }( {\bf k } , k_{ z } )
=
\pm \sqrt{ | \Delta |^{ 2 } + \left( E_{  \mbox{\scriptsize{Gr}} } - \mu \right)^{ 2 } }
\, ,
\label{EqDispersionGraphiteSC}
\end{equation}
with $ E_{  \mbox{\scriptsize{Gr}} }  $ given by Eq.~(\ref{EqDispersionGraphite}).

Therefore, the self-consistent equation for the superconducting gap becomes
\begin{eqnarray}
\frac{2}{\lambda }
&
=
&
\frac{ 1 }{ 2 }
\sum_{ j = 1}^{ 4 }
\int_{ - \frac{ \pi }{ d }  }^{ \frac{ \pi }{ d } } \frac{ d k_{ z } }{ 2 \pi d }
\,
\int \frac{ d^{ 2 } k }{ \left( 2 \pi \right)^{ 2 } }
\nonumber \\
& &
\hspace{1.25cm}
\frac{ 1  } { E_{ + }( {\bf k } , k_{ z } )  }
\tanh
	\left[
	         \frac{ \beta }{ 2 } E_{ + }( {\bf k } , k_{ z } )
	\right]
\, ,
\label{EqDVeff}
\end{eqnarray}
where the four values of $ E_{ + }( {\bf k } , k_{ z } ) $,
labeled by the index $ j $ in the above expression,
are given by $ E_{  \mbox{\scriptsize{Gr}} } $ in Eq.~(\ref{EqDispersionGraphite}),
in analogy to the discussions in the previous sections.

We calculate the critical temperature from Eq.~(\ref{EqDVeff})
and our results are compared with $ T_{ c } $ obtained for graphene bilayer,
from the former section.
Our results are shown in Fig.~\ref{FigTc_X_mu_Graphite}
and we see that there is not an enhancement of $ T_{ c } $ for every range of chemical potential for a given value of $ t_{ \bot } $.
However, given $ t_{ \bot }$,
we always find that the critical temperature for graphite is bigger than the $ T_{ c } $ obtained for graphene bilayer
for small values of the chemical potential,
what demonstrates that the first neighbors hopping between adjacent sheets favors the superconductivity in the system.

\section{Conclusions}
\label{Conclusions}

In conclusion, in the present paper we have derived the effective potential
for a stack of graphene layers with a hopping between adjacent sheets
and an on-site attractive interaction between electrons in a mean-field approximation.

For a single layer or two adjacent coupled layers of graphene,
a remarkable result was obtained for the superconducting critical temperature
as a function of the chemical potential:
it displays a dome-shaped curve, as experimentally observed in several compounds,
like 1111 pnictides and cuprate superconductors.
This result suggests that Dirac fermions may play a relevant role in the description of cuprates and iron pnictides,
which shall be object of further investigation.
Indeed, a dome-like structure of the superconducting phase
is in agreement with previous results
for strongly interacting two-dimensional Dirac fermions~\cite{Smith2009,Nunes2005}.
As pointed out in~\cite{Smith2009},
our results can also be experimentally realized with ultracold atoms in a two-dimensional optical square lattice.

Finally, considering a minimal model for graphite, taking into account only the
tunneling amplitudes between the nearest sites in the plane and out of the plane~\cite{Pershoguba2010},
we have compared the superconducting critical temperature for graphite and graphene bilayer.
We have seen that the $ T_{ c } $ calculated for graphite is bigger than the one for graphene bilayer for a small value of $ \mu $,
what might explain why intrinsic superconductivity is observed in HOPG.

\begin{acknowledgments}
This work has been supported in part by CNPq, FAPEMIG and FAPERJ.
We would like to thank N. M. R. Peres, H. Caldas, and A. H. Castro Neto 
for discussions on related matters.
\end{acknowledgments}



\bibliography{apssamp}

\begin{thebibliography}{400}

\bibitem{Neto_RMP09}
A. H. Castro Neto et al.,
Rev. Mod. Phys. {\bf 81}, 109 (2009).

\bibitem{Csanyi2005}
Cs\'anyi G et al.,
Nat. Phys. {\bf 1}, 42 (2005).

\bibitem{Weller2005}
T. E. Weller et al.,
Nat. Phys. {\bf 1}, 39 (2005).

\bibitem{Emery2005}
N. Emery et al.
Phys. Rev. Lett. {\bf 95}, 087003 (2005).

\bibitem{Belash2002}
I. T. Belash et al.,
Synt. Metals {\bf 36}, 283 (2002).

\bibitem{Hannay1965}
N. B. Hannay et al.,
Phys. Rev. Lett. {\bf 14}, 225 (1965).

\bibitem{Gunnarson1997}
O. Gunnarsson,
Rev. Mod. Phys. {\bf 69}, 575 (1997).

\bibitem{Kopolevich2007}
Y. Kopolevich,
J. Low Temp. Phys. {\bf 119}, 691 (2000);
Y. Kopelevich et al.,
Physics of the Solid State {\bf 41}, 1959 (1999)
[Fizika Tverd. Tela (St. Petersburg) 41 (1999) 2135].

\bibitem{Esquinazi2008}
P. Esquinazi et al.,
Phys. Rev. B {\bf 78}, 134516 (2008).

\bibitem{Savini2010}
G. Savini, A. C. Ferrari and F. Giustino
Phys. Rev. Lett. {\bf 105}, 037002  (2010).

\bibitem{Meng2010}
Z. Y. Meng et al.,
Nature {\bf 464}, 847 (2010).

\bibitem{Pathak2010}
S. Pathak, V. B. Shenoy and G. Baskaran
Phys. Rev. B {\bf 81}, 085431 (2010).

\bibitem{Kopnin2008}
N. B. Kopnin and E. B. Sonin,
Phys. Rev. Lett. {\bf 100}, 246808 (2008).

\bibitem{Baskaran2002}
G. Baskaran
Phys. Rev. B {\bf 65}, 212505 (2002).

\bibitem{Jiang2008}
Y. Jiang et al.,
Phys. Rev. B {\bf 77}, 235420 (2008).

\bibitem{Black-Schaffer2007}
A. M. Black-Schaffer and S. Doniach,
Phys. Rev. B {\bf 75}, 134512 (2007).

\bibitem{Roy2010}
B. Roy B and I. F. Herbut
Phys. Rev. B {\bf 82}, 035429 (2010).

\bibitem{Heersche2007}
H. B. Heersche, P. Jarillo-Herrero, J. B. Oostinga, L. M. K. Vandersypen, and A. F. Morpurgo,
Nature {\bf 446}, 56 (2007).

\bibitem{Pellegrino2010}
F. M. D. Pellegrino, G. G. N. Angilella and R. Pucci,
Eur. Phys. J. B {\bf 76}, 469 (2010).

\bibitem{Khveshchenko2009}
D. V. Khveshchenko,
J. Phys.: Condens. Matter {\bf 21}, 075303 (2009).

\bibitem{Honerkamp2008}
C. Honerkamp,
Phes. Rev. Lett. {\bf 100}, 146404 (2008).

\bibitem{Zhao2007}
E. Zhao and A. Paramekanti,
Phys. Rev. Lett. {\bf 97}, 230404 (2007).

\bibitem{Uchoa2007}
B. Uchoa B and A. H. Castro Neto,
Phys. Rev. Lett. {\bf 98}, 146801 (2007).

\bibitem{Marino2006}
E. C. Marino and L. H. C. M. Nunes
Nuc. Phys, B {\bf 741} [FS] 404 (2006).

\bibitem{Marino2007}
E. C. Marino and L. H. C. M. Nunes
Nuc. Phys. B {\bf 769} [FS] 275 (2007).

\bibitem{Nuno_Book_2007}
Eduardo V. Castro et al.,
An Introduction to the Physics of Graphene Layers,
in {\it Strongly Correlated Systems, Coherence and Entanglement},
(World Scientific, 2007)

\bibitem{Kremer2007}
R. K. Kremer, J. S. Kim and A. Simon
Carbon Based Superconductors,
in {\it High Tc Superconductors and Related Transition Metal Oxides},
(Springer-Verlag, Berlin Heidelberg, 2007)

\bibitem{Nunes2010}
L. H. C. M. Nunes, R. L. S. Farias. E. C. Marino,
{\it Superconducting and excitonic quantum phase transitions in doped systems with Dirac electrons}
{\bf ref do cond-mat}

\bibitem{Fukushima2007}
K. Fukushima and K. Iida,
Phys. Rev. D {\bf 76}, 054004 (2007).

\bibitem{mw}
N. D. Mermin and H. Wagner,
Phys. Rev. Lett. {\bf 17}, 1133 (1966);
P. C. Hohenberg,
Phys. Rev. {\bf 158}, 383 (1967);
S.Coleman,
Commun. Math. Phys. {\bf 31}, 259 (1973).

\bibitem{kt}
V.L.Berezinskii, Zh. Eksp. Teor. Fiz. {\bf 59}, 907 (1970);
J.Kosterlitz and D.Thouless, J. Phys. C {\bf 6}, 1181 (1973).

\bibitem{babaev}
E. Babaev,
Phys. Lett. B {\bf 497}, 323 (2001).

\bibitem{Smith2009}
L.-K. Lim et al.,
Eur. Phys. Lett. {\bf 88}, 36001 (2009).

\bibitem{Nunes2005}
L. H. C. M. Nunes and E. C. Marino,
Physica B {\bf 378-380}, 704 (2006).

\bibitem{Kampf1994}
A.P. Kampf,
Phys. Rep. {\bf 249}, 219 (1994).

\bibitem{cuprates}
I. Affleck and J. B. Marston
Phys. Rev. B  {\bf 37}, 3774 (1988);
Phys. Rev. B {\bf 39}, 11 538 (1989);
X-G. Wen and P. A. Lee
Phys. Rev. Lett. {\bf 76}, 503 (1996).

\bibitem{pnic1}
Y. Kamihara  et al.,
J. Am. Chem. Soc. {\bf 130}, 3296  (2008).

\bibitem{pnic2}
M. Rotter, M. Tegel M and D. Johrendt,
Phys. Rev. Lett. {\bf101}, 107006 (2008).

\bibitem{Direlpnic}
P. Richard et al.,
Phys. Rev. Lett. {\bf 104}, 137001 (2010).

\bibitem{Direlpnic1}
C. M. S. da Concei\c c\~ao, M. B. Silva Neto and E. C. Marino,
Phys. Rev. Lett. {\bf 106}, 117002 (2011).

\bibitem{Pershoguba2010}
S. S. Pershoguba and V. M. Yakovenko,
Phys. Rev. B {\bf 82}, 205408 (2010).

\end{thebibliography}


\end{document}